\documentclass[trackchanges]{aastex701}

\newcommand{\kms}{km s$^{-1}$}
\newcommand{\g}[1]{\color{blue}{#1}} 
\newcommand{\p}[1]{{\color{magenta}{#1}}}

\begin{document}

\title{Elongation of a Solar Filament and its Three-Dimensional Numerical Reconstruction for Magnetic Structures}

\correspondingauthor{Jinhan Guo}
\email{jinhan.guo@nju.edu.cn}

\author[orcid=0009-0003-3193-7496]{Garima Karki} 
\affiliation{Department of Physics, DSB Campus, Kumaun University, Nainital -- 263001, India}
\email{garimakarki31@gmail.com}

\author[orcid=0000-0002-4205-5566]{Jinhan Guo*}
\affiliation{School of Astronomy and Space Science and Key Laboratory of Modern Astronomy and Astrophysics, Nanjing University, Nanjing 210023, People’s Republic of China}
\affiliation{Centre for mathematical Plasma Astrophysics, Dept. of Mathematics, KU Leuven, 3001 Leuven, Belgium}
\email{jinhan.guo@nju.edu.cn}

\author[orcid=0000-0003-3364-9183]{Brigitte Schmieder}
\affiliation{Centre for mathematical Plasma Astrophysics, Dept. of Mathematics, KU Leuven, 3001 Leuven, Belgium}
\affiliation{LIRA, Observatoire de Paris, Universit\'e PSL, UMR8254 (CNRS), Sorbonne Universit\'e, Universit\'e Paris Cit\'e, 5 place Jules Janssen, 92195 Meudon, France}
\affiliation{SUPA, School of Physics \& Astronomy, University of Glasgow, Glasgow G12 8QQ, UK}
\affiliation{LUNEX EMMESI COSPAR-PEX Eurospacehub, Kapteyn straat 1, Noordwijk2201 BB Netherlands}
\email{brigitte.schmieder@obspm.fr}

\author[orcid=0000-0002-3518-5856]{Ramesh Chandra}
\affiliation{Department of Physics, DSB Campus, Kumaun University, Nainital -- 263001, India}
\email{rchandra.ntl@gmail.com}

\author[orcid=0000-0001-8215-6532]{Pascal D\'emoulin}
\affiliation{LIRA, Observatoire de Paris, Universit\'e PSL, UMR8254 (CNRS), Sorbonne Universit\'e, Universit\'e Paris Cit\'e, 5 place Jules Janssen, 92195 Meudon, France}
\email{pascal.demoulin@obspm.fr}
\author[orcid=0000-0002-1743-0651]{Stefaan Poedts}
\affiliation{Centre for mathematical Plasma Astrophysics, Dept. of Mathematics, KU Leuven, 3001 Leuven, Belgium}
\affiliation{Institute of Physics, University of Maria Curie-Skłodowska, Lublin, Poland}
\email{stefaan.poedts@kuleuven.be}

\author{Bernard Gelly}
\affiliation{THEMIS, Canary Islands, ES}
\email{bgelly@themis.iac.es}

\begin{abstract}

Quiescent filaments are prominent features of the solar atmosphere, and their evolution reflects the coronal magnetic field’s response to photospheric magnetic activity. Here, we report on a quiescent filament observed from 2023 September 28–29, aiming to understand how the magnetic configuration shapes its feet and drives its extension. For this purpose, high-resolution spectral data in H$\alpha$ and Mg II  k are used from the T\'elescope H\'eliographique pour l’Etude du Magn\'etisme et des Instabilit\'es Solaires (THEMIS) and the Interface Region Imaging Spectrograph (IRIS), respectively. To track changes in the filament, we utilise long-term data from the Atmospheric Imaging Assembly (AIA) on the Solar Dynamics Observatory (SDO) and from the Global Oscillation Network Group (GONG). We analyse the longitudinal magnetic field in the photosphere using the Solar Optical Telescope (SOT) onboard Hinode, as well as SDO/Helioseismic and Magnetic Imager (HMI) data. In addition to this, we use GONG H$\alpha$ data to analyze the longitudinal oscillations in the filament. Observations show that parasitic polarities and canceling flux play a key role in forming and reorganizing the filament feet and in lengthening the filament. A 3D MHD reconstruction using vector magnetograms reveals that its magnetic configuration evolves into a full flux rope (FR), whose extension on the second day matches the observed filament growth. The FR is separated from the surrounding nearly potential field by quasi-separatrix layers, which in turn are separated by current layers. They get more organized around the FR as it is growing up. Moreover, the longitudinal oscillations in the extended filament are attributed to heating from flux cancellation in underlying bright points.
\end{abstract}


\section{Introduction}
\label{Intro}

\subsection{Formation of filaments}
Solar filaments (known as prominences at the solar limb) are relatively cold (10$^{4}$ K) and dense plasma embedded in the hot corona \citep[10$^{6}$ K,][]{Tandberg1974,Tandberg1995,Labrosse2010,Mackay2010}. Filament fine structures are composed  of thin and dark fibrils \citep{Lin2005,Karki2025}. The main thin structure running along the filament is called the spine. This is typically understood as the top part of the filament.  Lateral extensions of fibrils are typically present on both sides of the filament, forming a global structure known as feet or barbs. 
All the filament fibrils are slightly inclined with respect to the polarity inversion line (PIL). In magnetic structures, including magnetic dips such as the flux rope (FR) and the sheared arcade, the dense plasma is supported against gravity by the Lorentz force. 

It is important to understand filament formation because filament eruptions are frequently associated with the onset of coronal mass ejections (CMEs). Filament first starts to rise slowly \citep{Chandra2021, Devi2021, Cheng2020}, then evolves to erupt through kink, torus instability and fast magnetic reconnection \citep{Torok2006, Kliem2006, Aulanier2010, Jiang2021_f, Xing2024}. 
Depending on the environment of the FR, it can escape and accelerate particles in open magnetic field lines or be confined if the magnetic field strength is too strong above the FR \citep{Amari2018}, or if the Lorentz force leads to an increase in the electric current, forcing the FR to rotate and ultimately stop its rising phase \citep{Zhang2024, Guojh2024}.
During an eruption, the FR is expelled and evolves into a CME.  When a CME carrying a negative (southward) $B_z$ component reaches Earth's environment, it induces magnetic reconnection within Earth's intricate magnetosphere, triggering geomagnetic storms.

There are various models for forming a filament, such as injection, levitation, and (or) evaporation-condensation \citep{Mackay2010, Zhou2025, Keppens2025}. The injection model is more relevant for active region (AR) filaments, and the levitation model is relatively well accepted when chromosphere fibrils gather together and rise. Nevertheless, the evaporation--condensation model is, to date, the best accepted. Many hydrodynamic (HD) or magnetohydrodynamic (MHD) simulations of filament formation have been developed around this idea \citep{Karpen2003,  Karpen2006, Xia2011, Xia2014, Zhou2014, Zhou2020, Guo2021_2Dip,Liakh2025}. Extensive studies have been performed theoretically on the dips present in an FR to characterize parameters such as the dip depth, the dip extension along the main axis, and the number of dips \citep{DeVore2005,Zhou2017,Guo2021_2Dip,Guo2022_prom}.  It was concluded that magnetically connected threads due to multiple dips are more likely to exist in quiescent filaments than in AR filaments \citep{Guo2021_2Dip}.

\subsection{Magnetic configuration of filaments}
Filaments are classified as AR filaments when present inside an AR, quiescent filaments when located in the quiet Sun (typically at high latitudes), and intermediate filaments when they lie between two ARs or have one foot rooted within an AR.
They are commonly modeled as sheared arcades or twisted flux ropes \citep{Aulanier1998_D, Chen2014, Ouyang2017}, with approximately 90\% of filaments supported by twisted flux ropes \citep{Ouyang2017}.
In particular, even a filament might be supported by an FR in one segment and by a sheared arcade \citep{Guoy2010, Liht2025}. It has been demonstrated that large twists are present in quiescent filaments, leading to the formation of short threads, while in AR filaments, there are longer threads \citep{Guo2021_2Dip, Guo2022_prom}. Nowadays, data-driven simulations demonstrate the formation of FRs as the AR evolves with shearing motions and moving polarities \citep{Jiang2021_f, Guo2024_data, Schmieder2024, Wangws2025, Liht2025}. The evolution of photospheric magnetograms is crucial for understanding filament formation and evolution.

\subsection{Magnetic Barbs $/$ Feet}
\label{subS:Magnetic_barbs} 
Filaments are located between two inverse photospheric magnetic polarities separated by the polarity inversion line (PIL).
Viewed on the limb, filaments/prominences are attached to the disk by regularly spaced legs or feet. 
Viewed on the disk, the feet are also called barbs \citep{Martin1992}. In this paper, we will preferentially use the word feet. Commonly, feet are related to parasitic magnetic polarities, which have the opposite sign from the dominant magnetic polarity on each side of the PIL. Feet shape and existence depend on the location and magnetic flux of the parasitic polarities \citep{Aulanier2000}. Feet are located on magnetic field lines that bend towards the photosphere, where they are attracted by parasitic polarities. Then, the magnetic field lines present dips, away from the main filament body, where cool plasma can be trapped  \citep{Aulanier1998_D, Aulanier1999}.  

Some magnetic field lines with dips extend to the photosphere in the filament channel, forming feet. Plasma in feet does not drain back to the photosphere, unlike at the filament ends \citep[e.g.,][]{Schmieder1985,Guo2021_2Dip}.    Any motion of photospheric granules and supergranules affects the dynamics of the filament. Supergranules are well identified in the chromosphere \citep{Schmieder2014,Zhou2021}.  In \citet{Schmieder2024},  converging flows were identified around  H$\alpha$ filament feet and at the edges of the EUV filament channel, which is wider than the H$\alpha$ channel \citep{Aulanier2002}. The horizontal photospheric velocity may reach 1 \kms. In some locations, horizontal flows crossing the channel are observed, which eventually indicate large-scale shearing \citep{Schmieder2014}. 

The existence of parasitic polarities forming barbs or feet remains controversial \citep{vanBallegooijen2004,Liu_Xia2022,Chen2025_Xia}. The existence of dips cannot be directly proven by observation. 
%This split topology is due to strong network polarities on the edge of the filament channel as it was  shown in 
\citet{Dudik2008} studied  the topological departures from translational invariance along a filament. They calculated the coronal magnetic field from a ``linear magnetohydrostatic'' extrapolation of a composite magnetogram and detailed the shape of the dips corresponding to a long observed filament and their feet. They showed the importance of the network at the border of  filament for the  split topology in  the filament.
Other types of barbs/feet have been recently described as follows: dynamic barbs due to the longitudinal oscillations of some threads \citep{Ouyang2020}, and barbs due to the indented threads \citep{Chen2020, Guo2022_prom}. These barbs do not correspond to any prominence footpoint. Below, we use the term `barb' to refer only to dynamic lateral extensions, while `feet' is reserved for the most stationary lateral/downward extensions, which still evolve but on long timescales (hours, days) relative to the photospheric field.

Filament feet and spines %in ARs 
resemble an accumulation of parallel, short threads \citep{Karki2025}. These short threads can be formed by the cooling of plasma in coronal loops \citep{Karpen2006,Luna2012}. They can also form by merging adjacent chromospheric fibrils due to strong shear \citep{Joshi2022}. In that paper, they conjectured that the squeezed fibrils were rising and forming a longer filament. 
Such filament formation has been proposed recently using MHD numerical simulations of supergranules \citep{Chen2025_Xia}. 
In this case, prominence material, including feet, remains in the dip of the helical field lines, which are piled up vertically from the photosphere to the spine.

\subsection{Chirality}
Feet/barbs indicate the chirality of the filament. If they are inclined like the highway exits in the European mainland, they are right-bearing filaments \citep{Martin1992,Tandberg1995,Lopez2006,Chen2014,Chen2020}. In contrast, if the feet/barbs organisation is the mirror view, they are left-bearing filaments.  
Another filament chirality is defined by the direction of the axial magnetic field. For an observer located on the positive-polarity side of the PIL, the configuration is called dextral if the axial magnetic field points to the right, and sinistral if it points towards the left. \cite{Martin1998} found a correlation between these two types of chirality, i.e., left-bearing/right-bearing feet correspond to sinistral/dextral filaments. For FR models, such correlation naturally arises from the positive/negative magnetic helicity of the FR \citep{Aulanier1998_D}. 

Magnetic arcade above the FR is expected to have the same helicity sign (since the FR is formed by the reconnection of part of this arcade). Then, the coronal arcade above the filament is expected to be right- or left-skewed, respectively, as typically observed \citep[see][for a more comprehensive review]{Mackay2010}.   
The ARs with a preponderance of positive/negative helicity are in the southern/northern hemisphere \citep{Ouyang2017}. Moreover, cases have been reported where dextral and sinistral filaments coexist in the same hemisphere \citep{Chandra2010, Ouyang2017}.  

\subsection{Large amplitude Oscillations}
Observations of large-amplitude oscillations in filaments, defined by a velocity greater than 10~\kms,  are frequent nowadays \citep[see the review of][]{Arregui2018}. The detection of these oscillations is possible because of the regular, long-time interval observations provided by space- and ground-based instruments. The triggers are various \citep{Luna2018}. They can be due to Moreton or EUV waves \citep{Asai2012, Devi2022}, shock waves \citep{Shen2014,Jercic2022}, nearby jets \citep{Joshi2023}, subflares and flares or related to the eruptive phase of a filament \citep{Bocchialini2011}.  \citet{Luna2018} analyzed the oscillations detected in 196 filaments observed with GONG. Nearly half of them have a large amplitude velocity  (v $> $ 10 \kms). 
The flow direction makes an angle of about 20 degrees relative to the filament spine.  Considering that fine structures of filament trace the local supporting magnetic field lines, they deduced that the flow was along the strands of the filament and, therefore, it should be longitudinal oscillations. The average period was 58 $\pm\;$15~min.

The prominence oscillatory periods depend on height, with periods increasing by up to 10-15 min for a rise in height of approximately 30-40 Mm \citep{Hershaw2011}. 
Damping of the oscillations is also observed in many studies \citep{Bi2014}. 
In their observations, the oscillation period increased from 60 min to 90 min after two oscillation periods. They suggest that the curvature of magnetic fields supporting the filament becomes flatter during the evolution phase, leading to longer oscillation periods. 

Transverse and longitudinal oscillations could coexist in many cases \citep{Mazumder2020, Tan2023}. The observed velocity profiles are typically fitted with an exponentially decaying sine function. This allows the estimation of wave periods, damping times, and initial amplitudes.  

The theoretical model of \citet{Luna2012} for the large amplitude oscillations suggests the following scenario: after an energetic event (a subflare, for instance) occurring close to a filament, the injected energy evaporates plasma at the flux tube footpoint closest to the energetic event. Then, the upward flow of hot plasma pushes the cold plasma condensations (threads) located in the dips of the magnetic structure, and longitudinal oscillations start.
The oscillations are gravity-driven, with the restoring force being the projected solar gravity along magnetic field lines. 
More generally, several theoretical models are based on pendulum models with adaptations that take into account the source and possible wave reflections \citep{Luna2012, Jercic2022}. The damping time depends on the curvature of the dips and the altitude. It may also be affected by multi-threads that reconnect \citep{Zhou2017}.

\begin{figure}    %%%%%%%%%%%%%%%%%% FIGURE 1 
\centerline{\includegraphics[width=1.\textwidth,clip=]{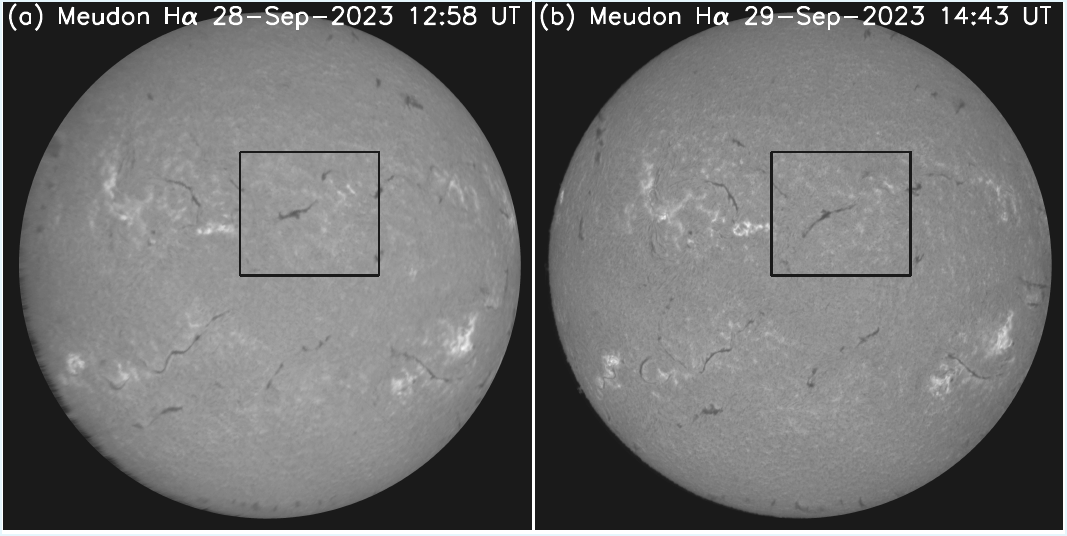}}
    \caption{Panels (a) and (b) show the full disk H$\alpha$ image observed by Meudon spectroheliograph on 28 and 29 September 2023, respectively. The black box indicates the filament observed by THEMIS and IRIS during the 2023 campaign. 
    We note the change in shape of the east end of the filament on September 29.}
\label{Meudon}
\end{figure}

\begin{figure}    %%%%%%%%%%%%%%%%%% FIGURE 2
\centerline{\includegraphics[width=1.\textwidth,clip=]{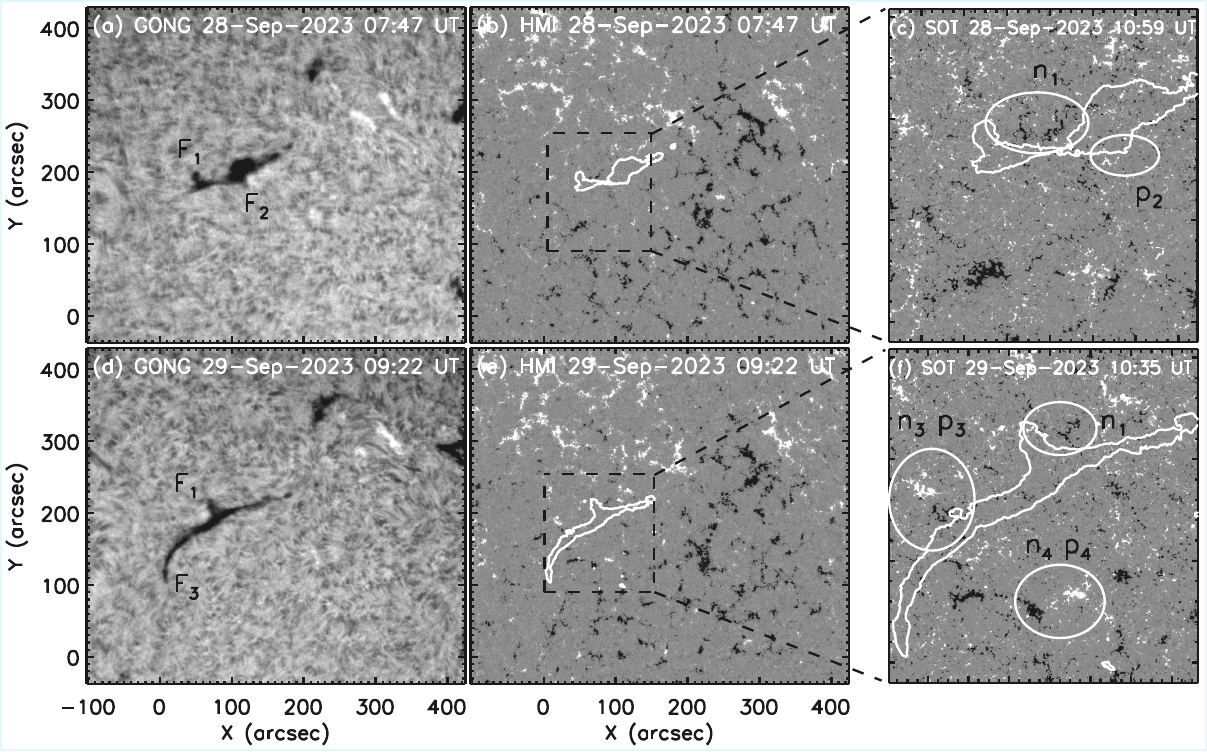}}
\caption{Zoom on the filament observed with GONG H$\alpha$ (panels a and d), the longitudinal magnetic field (HMI in panels b and e) and a local zoom of Hinode/SOT with levels $\pm$ 100 G (in panels c and f) for September 28 and September 29. The two FOVs of the filament are aligned with the FOV of 29 Sept at 11:00 UT.
We note that the global pattern of the network polarities of HMI is well co-aligned.  
The FOV of SOT is represented by the dashed black box in HMI magnetograms. 
F$_1$, F$_2$, and F$_3$ are feet of the filament. On September 28, two feet, F$_1$ and F$_2$, are observed, while on September 29, feet F$_1$ and the extension F$_3$ are present. 
In the SOT maps, ovals indicate the parasitic polarities (n$_1$, p$_2$) and cancelling flux regions (n$_3$ p$_3$, n$_4$ p$_4$). More details on the observed data sets, including the observed wavelengths, are provided in the Appendix~\ref{appendix}.
An animation of panels a and d is available, starting on 27 September 2023 at 23:59 UT and ending on 29 September 2023 at 23:58 UT. The real-time duration of the animation is 3 min 12 s. (An animation of this figure is available in the online article.)}
\label{GONG_HMI_SOT}
\end{figure}

\begin{figure}    %%%%%%%%%%%%%%%%%% FIGURE 3
\centerline{\includegraphics[width=1.\textwidth,clip=]{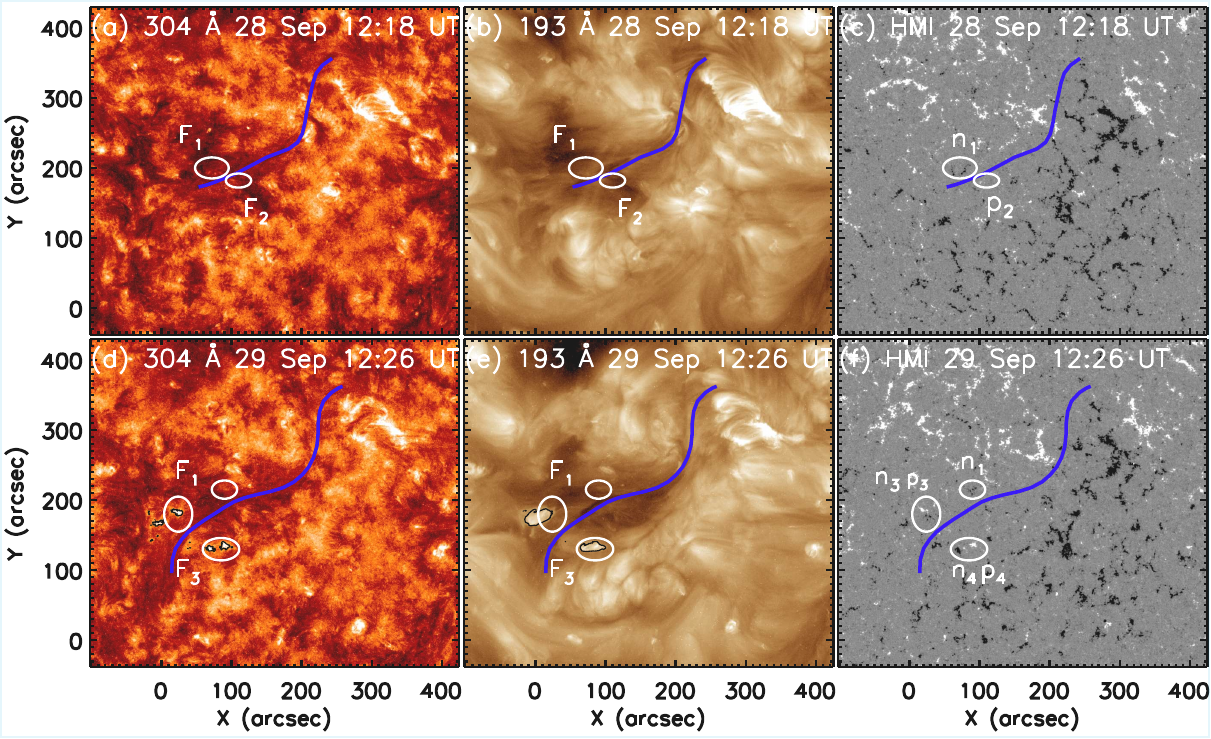}}
\caption{Region of the filament observed in AIA 304 and 193 \AA\ on the 28$^{th}$ and 29$^{th}$ September 2023 (panels a, d and b, e, respectively). The FOVs are aligned on the FOV of September 29 at 11:00 UT. HMI magnetograms with levels $\pm$ 100 G of the filament region for September 28 and 29 are shown in panels (c) and (f), respectively. The blue curve represents the filament spine traced along the dark part of the filament from AIA 193 \AA\ images for the 28$^{th}$ and 29$^{th}$ of September, respectively. 
The feet F$_1$, F$_2$, and the extension F$_3$ are indicated in panels a, b, d, and e. The polarities n$_1$, p$_2$ are indicated in panel c, and n$_3$, p$_3$, n$_4$, p$_4$ in panel f. The ellipses marked in the AIA filter images correspond to those in the HMI magnetograms, denoting the parasitic polarities (n$_1$, p$_2$) and the bipoles (n$_3$–p$_3$ and n$_4$–p$_4$). The black contours in panels d and e outline the observed EUV bright points. An animation of this figure is available, starting at 00:00 UT on 28 September and ending at 23:57 UT on 29 September. 
The real-time duration of the animation is 32 s.(An animation of this figure is available in the online article.)}  
\label{AIA_HMI}
\end{figure}

\begin{figure}    %%%%%%%%%%%%%%%%%% FIGURE 4
\centerline{\includegraphics[width=1.\textwidth,clip=]{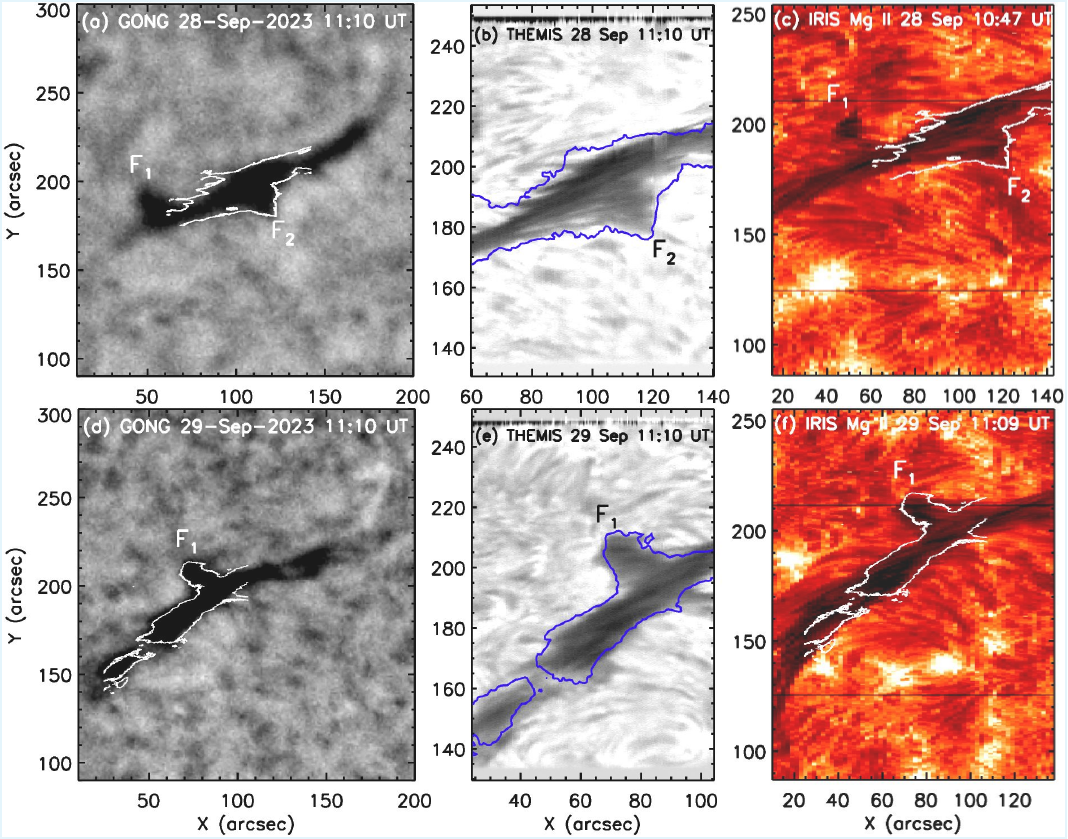}}
\caption{Zoomed view of the filament observed by GONG (panel a) in H$\alpha$, and the slit-reconstructed
images from the spectra in H$\alpha$ line center observed by THEMIS (panel b), and in Mg II k line center (2796.4 \AA) by IRIS
(panel c), for September 28 and in d, e, f for September 29. The contours (blue) of the filament observed in GONG are overlaid on the THEMIS image, and the filament contours from THEMIS are overlaid in white over GONG and IRIS images to show the THEMIS FOV, which is focused on F$_2$ on September 28 and on F$_1$ on September 29. These images are aligned with the X coordinate at 11:00 UT on September 29. Fine structures in the filament are visible in THEMIS and IRIS maps.}
\label{THEMIS_IRIS}
\end{figure}

\begin{figure}    %%%%%%%%%%%%%%%%%% FIGURE 5
\centerline{\includegraphics[width=1.\textwidth,clip=]{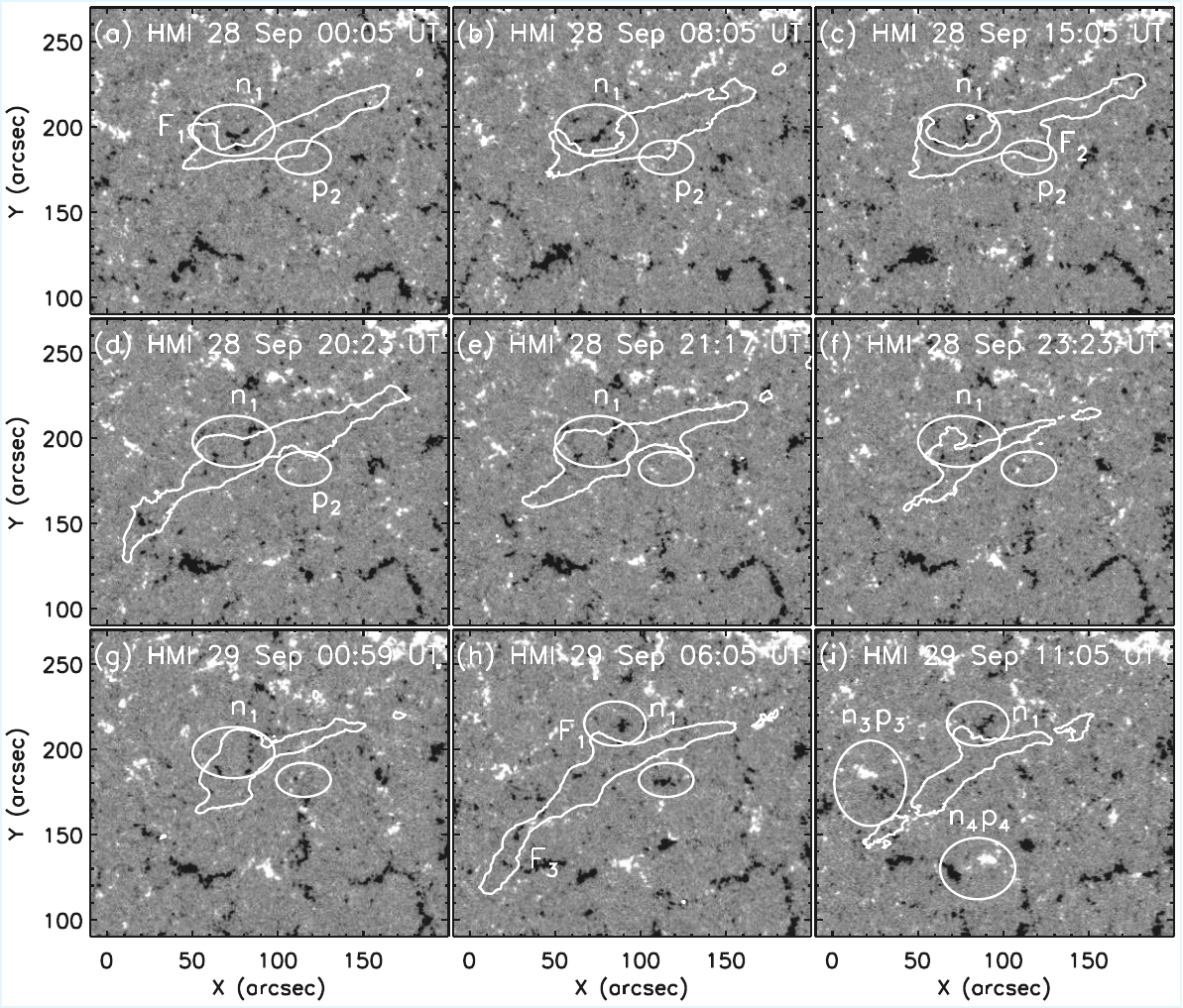}}
\caption{
Filament contours and magnetic field evolution. The magnetograms are displayed with magnetic field levels of $\pm$50 G for the positive and negative polarities, respectively. The white contours in each panel are the filament contours observed in GONG corresponding to the HMI magnetogram time. 
Panel a correspond to the initial phase of the formation of foot F$_2$ located near the very small positive polarity p$_2$. 
Panel b corresponds to the $1^{st}$ maximum of polarity p$_2$ at around 08 UT, shown by the left red arrow in Figure~\ref{HMI_time} b. 
Panel c shows a slight decrease in p$_2$. 
Panel d is when F$_2$ has disappeared. 
Panel e is when F$_2$ reappears around 21 UT.
Panel f corresponds to the $2^{nd}$ maximum of p$_2$. We also observe a small negative polarity within p$_2$ and the disappearance of F$_2$. 
In panel g, the negative polarity within p$_2$ is stronger, and none of the filament feet are clearly identifiable. 
In panel h, a strong negative polarity has emerged at the location of p$_2$, which corresponds to the maximum of the blue curve in Figure~\ref{HMI_time} b at around  06 UT on September 29. 
In panels h and i, we observe two bipoles n$_3$ p$_3$ and n$_4$ p$_4$ (ellipses in panel i) and the extension of the filament up to F$_3$. An animation of this figure is attached. The animation starts at 23:59 UT on 27 September and ends at 23:53 UT on 29 September. The animation's real-time duration is 1 min 20 s. (An animation of this figure is available in the online article.)}

\label{GONG_HMI}
\end{figure}

\begin{figure}    %%%%%%%%%%%%%%%%%% FIGURE 6
\centerline{\includegraphics[width=1.\textwidth,clip=]{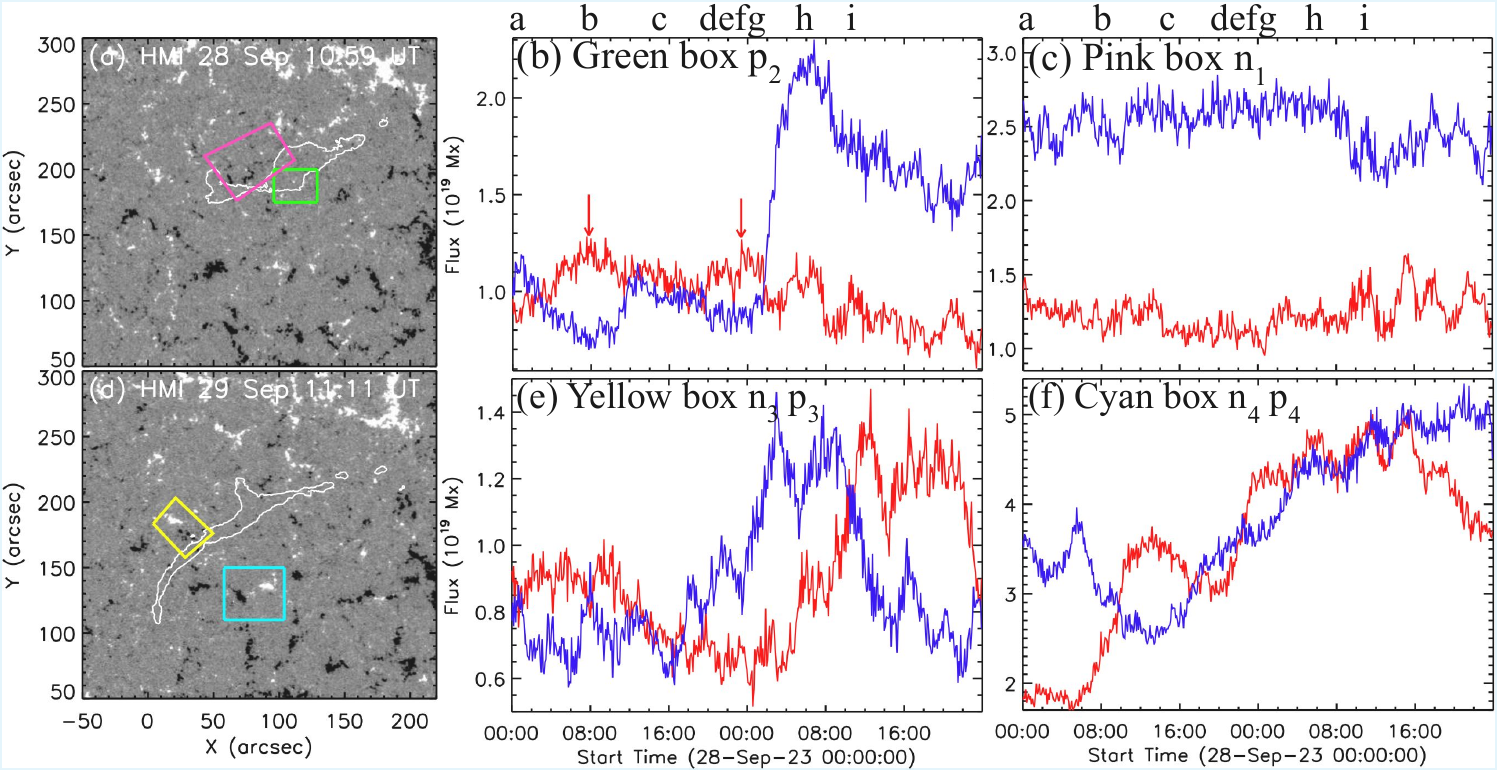}}
   \caption{HMI line-of-sight magnetograms  (with levels $\pm$ 70 G) of the filament channel for September 28 and 29 in panels a and d, respectively. The filament contours observed in GONG for the corresponding times are overlaid in panels a and d.
   The temporal evolution of the magnetic flux is shown in panels b (green box), c (pink box), e (yellow box), and f (cyan box), with the boxes in panels a and d. The green box is located over polarity p$_2$ and the panel labels of Figure~\ref{GONG_HMI} are added on the panel top. The red and blue curves correspond to positive and negative polarity, respectively. The red arrows in panel b correspond to the maximum of the positive parasitic polarity p$_2$. An animation of panels a and d is available, starting on 27 September at 23:59 UT and ending at 23:53 UT on 29 September. The animation's real-time duration is 1 min 20 s. (An animation of this figure is available in the online article.)}
\label{HMI_time}
\end{figure}

\subsection{Aim and outline of the paper}
The goal of this paper is to understand the formation of the feet and the elongation of a filament observed during two days. We used multi-wavelength observations of a quiescent filament obtained with ground-based and space-based instruments (Section \ref{Instrument} and Appendix). In Section \ref{obs}, we describe the morphology of the filament and its magnetic environment. In Section \ref{simu}, we reconstruct their 3D magnetic structures with the non-linear-force-free-field (NLFFF) extrapolation. Section \ref{con} presents the conclusion.

\section{Filament observations}
\label{obs}
\subsection{Alignment of the data}
\label{Instrument}

The filament that we choose to analyse is a quiescent filament on the solar disk observed for 2 days, 28-29 September 2023, using the following space- and ground-based observatories (see the full description of the instruments in the Appendix): the Solar Dynamics Observatory \citep[SDO;][]{Pesnell2012}, the Interface Region Imaging Spectrograph \citep[IRIS;][]{DePontieu2014}, the Hinode/Solar Optical Telescope \citep[Hinode/SOT;][]{Tsuneta2008}, the Global Oscillation Network Group \citep[GONG;][]{Harvey1996}, the T\'elescope H\'eliographique pour l’Etude du Magn\'etisme et des Instabilit\'es Solaires \citep[THEMIS;][]{Mein1985}.
We focus on the evolution of this filament because it was the target of the THEMIS and IRIS observation campaign and was observed with these two high-spatial resolution instruments \citep{Schmieder2025,DePontieu2014}.

The THEMIS observations acquired with a small pixel size of 0.06$''$ along the slit have been described for September 29 in a previous paper \citep{Karki2025}. The data from September 28 were acquired in a similar manner between 08:23 UT and 11:19 UT with the scans numbered t01 to t20. They consist of observations of different sections of the filament taken irregularly over time.
%due to the small field of view (FOV), and, moreover, they lack an automatic cadence. 
Some of them (e.g.,  t019; THEMIS observation file at 11:10 UT) were carried out with accumulation ($10 \times 200\;$ms = 2\;s), a step size of 0.5$''$, covering a span of $\pm\ 40''$, and a slit length ranging between 110 -- 120$''$. Some of the spectra are very good during each scan, particularly between 10:10 UT and 11:19 UT (three images: 10:38 UT, 11:10 UT, and 11:19 UT), but the reconstructed images are not homogeneous due to the rapid changes in seeing, which are  corrected by the adaptive optics.

The Meudon spectroheliograph in the H$\alpha$ wavelength is  used to study the global structure of the filament. The observations are obtained by scanning the full disk in one minute  with an image scale of 1$''$.1/pixel.

All data sets for September 28 and 29 are aligned to a common reference time (September 29, 2023, at 11:00:00 UT) to facilitate a comparative analysis between the two days with range [-102, 425]  arcsec in x and [-38, 430] arcsec in y.
Observations from SDO, GONG, and Meudon were corrected for the solar differential rotation using the drot\_map routine available in SolarSoft system (SSW). Residual misalignments between SDO and GONG were manually corrected by comparing the centroids of sunspots visible in HMI continuum images and GONG images simultaneously. 
IRIS shows good alignment with AIA, with minor misalignments corrected by comparing bright points in the IRIS slit-jaw images with AIA 304 \AA\ images. Lastly, THEMIS was aligned with GONG and IRIS by comparing filament contours. 
The SOT/SP longitudinal magnetic field magnetograms were manually aligned with HMI data by comparing the magnetic polarities observed in both datasets. We also use the HMI vector magnetograms for the 3D reconstruction of filament magnetic structures.

\subsection{Morphology}
A filament near the central meridian was the target of the THEMIS campaign for two days on September 28 (N17, E07) and September 29, 2023  (N17, W06).
The filament extends in the middle of the added black squares in Figure~\ref{Meudon}. A small, separated filament portion is also present in the north-west (still within the squares).

Figure~\ref{GONG_HMI_SOT} presents snapshots of the region, one for each day observed by GONG (panels a, d). 
The filament, located around the panel centers, is mostly oriented east-west at an angle of about 10$^{\circ}$, with respect to the solar equator on September 28.  Two feet are present on September 28 (F$_1$ and F$_2$), and only one (F$_1$) is present on September 29. On the second day, the eastern part of the filament extends and bends southward up to F$_3$.
Panels b and e of Figure~\ref{GONG_HMI_SOT} show snapshots of the HMI movie for these two days with levels $\pm$ 100 G.
The filament lies along the PIL between positive polarity in the north and negative polarity in the south, spanning a length of  $\approx$ 180-380 arcsec. %15-20 degrees in heliographic coordinates. 
It is a region of a network with identifiable supergranules.  

The campaign targeted the southern/main part of the filament, represented by the  white contour in Figure~\ref{GONG_HMI_SOT} panels (b, c, e, f). 
The presence of feet is correlated with the presence of parasitic polarities visible in the  SOT/SP magnetograms with levels $\pm$ 100 G (Figure~\ref{GONG_HMI_SOT} panels c and f). Foot F$_1$ is related to the parasitic polarity n$_1$, and foot F$_2$ is related to the positive polarity p$_2$. 
The filament extends up to F$_3$ between the two new bipoles (n$_3$ p$_3$, n$_4$ p$_4$) on September 29 (panel f).

Figure~\ref{AIA_HMI} compares the filament images observed on September 28 and 29 in AIA 304 \AA, AIA 193 \AA, and the corresponding HMI magnetograms. 
In the 304 \AA\ images, we identify a filament channel up to F$_1$ at 12:18 UT on September 28 (Figure~\ref{AIA_HMI} panel a). 
On September 29, the filament channel extends to the south and corresponds to the end F$_3$ (Figure~\ref{AIA_HMI} panel d).  On both sides of F$_3$ there are two bright, tiny regions that correspond to the bipoles n$_3$ p$_3$, and  n$_4$ p$_4$ (Figure~\ref{AIA_HMI} panel f). These bright points are also well visible in 193 \AA\ (Figure~\ref{AIA_HMI} panel e). We will discuss their role in Section \ref{extension}.

In 193 \AA\  the optical thickness of the cool material absorption is similar to the optical thickness in H$\alpha$ \citep{Anzer2005}. Therefore, the denser part of the filament appears as a strong dark line and corresponds to the region with the largest optical thickness in the filament in H$\alpha$ (panels b, e).  This line is overlaid on the HMI magnetograms (blue curve in all panels). 
We observe that the northwestern part of the filament exhibits a similar structure on both days, though it is slightly more curved on September 29.

\subsection{Evolution of feet F$_1$ and F$_2$ versus the evolution of the magnetic field}
\label{evolution}
The coordinated observations of THEMIS and IRIS were focused on September 28 on foot F$_2$ and on September 29 on foot F$_1$.
Figure~\ref{THEMIS_IRIS} shows the relationship of the observations of feet F$_1$ and F$_2$ with THEMIS and IRIS for the two days. The high spatial resolution of the instruments allows us to see the fine strands along the filament and in the feet. 
The adjacent threads have widths of the order of one arcsec. The fine structure makes a small angle with respect to the main axis (about 20 degrees). The foot F$_1$ has been well studied in H$\alpha$ and Mg II lines in a previous paper, which showed strong counterstreaming in the adjacent threads, with velocities of up to 20 \kms\ \citep{Karki2025}.   

The evolution of the feet F$_1$ and F$_2$ is presented in the GONG movie and with white contours in Figure~\ref{GONG_HMI}. The filament remains highly dynamic on both days. Moreover, the filament changes its shape drastically between 15 UT (1$^{st}$ day, panel c) and 06 UT (2$^{nd}$ day, panel h).
On the first day, the foot F$_1$ is clearly observed at 00 UT and is related to the negative polarity n$_1$ (panel a). F$_1$ gradually fades between 15  UT and 20 UT (panels c, d), from its initial position, and reappears slightly north-westward at around 23 UT (panel f). After that, F$_1$ changes its shape and size continuously, becoming more extended around 06 UT  on the next day (panel h), and remains in this state throughout the day. 
Next, the foot F$_2$ is well-defined between 00:00 UT and 15:00 UT on September 28 (panels a-c), then it becomes diffused and fades out. The foot F$_2$ reappears around 21 UT (panel e) and then disappears the next day.

The HMI movie of the two days shows the evolution of the magnetic pattern in the filament region.  
Figure~\ref{HMI_time} summarizes the time evolution of the magnetic field flux in boxes containing the parasitic polarities n$_1$, p$_2$, n$_3$ and p$_3$,  n$_4$ and p$_4$ related to F$_1$, F$_2$, F$_3$, respectively. For the flux computation within the boxes, all HMI magnetograms are aligned to a common reference time (September 29, 2023, at 11:00:00 UT) to compensate for solar rotation. The boxes used to compute the magnetic flux are defined on the aligned data to properly track the magnetic polarities.
The panel labels of Figure~\ref{GONG_HMI} have been added to the top of the panels of Figure~\ref{HMI_time} to link the magnetic flux evolution to the magnetogram evolution.

The polarity n$_1$ (pink box in panel a)  is present on both days and is moving slightly to the east. This is why we adopt an elongated box shape to follow n$_1$. We have noted previously that F$_1$ is also progressively moving slightly to the east. 
The magnetic flux measured in the corresponding box containing n$_1$ shows a nearly constant high negative flux for both days, around 2.6 $\times$ 10$^{19}$ Mx  (Figure~\ref{HMI_time} c).
The parallel evolution of n$_1$ and F$_1$ further illustrates the model of \citet{Aulanier1998_D} and \citet{Aulanier1999}, which shows that a filament foot is due to the presence of a parasitic polarity and evolves with it.

The formation of F$_2$ is related to the increase in the magnetic flux of p$_2$ with a maximum flux of $1.3 \times 10^{19}\;$Mx (see the left red arrow close to the red line in Figure~\ref{HMI_time} panel b), while the negative flux (blue line) is decreasing to $0.7 \times 10^{19}\;$Mx between 00:00-08:00 UT on September 28. Later on, F$_2$ reappears on September 28 at around 21 UT (Figure~\ref{GONG_HMI}, panel e), which can be explained by the increase in the positive polarity to $1.2 \times 10^{19}\;$Mx at around 20 UT (right red arrow in Figure~\ref{HMI_time} panel b).
The disappearance of F$_2$ on September 29 is associated with a decrease in the positive flux, while the negative flux remains steadily increasing. These related evolutions provide additional evidence for the link between filament foot and parasitic polarity.
Finally, we note the rapid increase of negative magnetic flux (blue line) in Figure~\ref{HMI_time} panel b after 06:00 UT. This corresponds to the entrance of a negative polarity in the selected green box, and it could also contribute to the disappearance of F$_2$.

\begin{figure}[!t] %%%%% FIGURE 7
    \centering
\includegraphics[width=\textwidth]{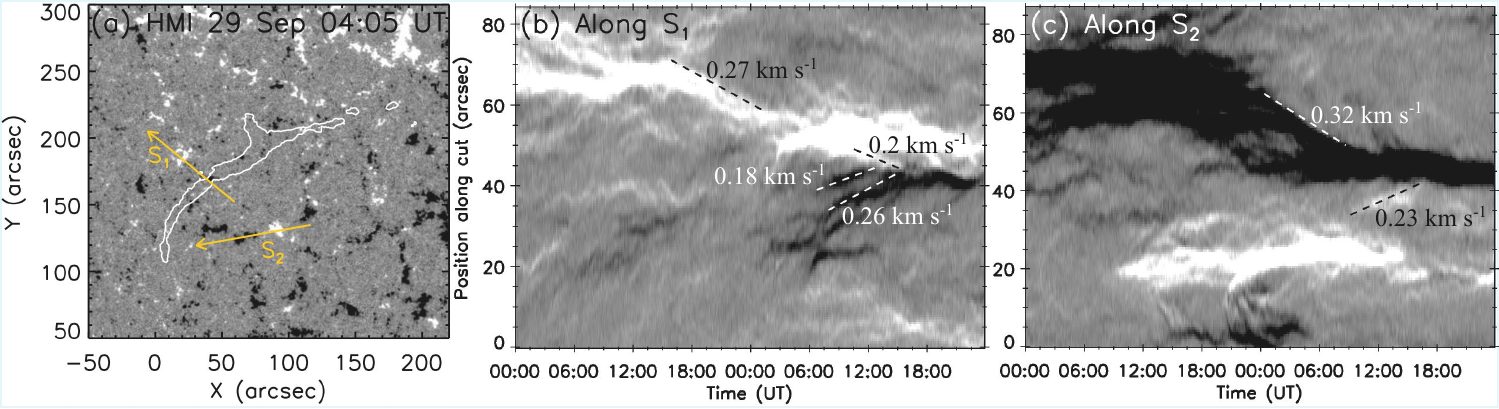}
    \caption{HMI magnetogram ($\pm$ 50 G) on 29 September 2023 with filament contours in white (panel a). S$_1$ and S$_2$ show the slit positions defined to follow the local motions of bipoles n$_3$ p$_3$ and n$_4$ p$_4$. 
    The slit width is 4.8$''$. Panels b and c show the time-distance diagrams for slits S$_1$ and S$_2$, respectively. Straight lines are defined to follow locally the polarity motions. Their slopes provide the indicated velocities.}    
    \label{hmi_ts}
\end{figure}

\subsection {Extension/elongation  of the filament to F$_3$}
\label{extension}

The filament extends to F$_3$ on September 29 (Figure~\ref{GONG_HMI_SOT} d). The extension F$_3$ and its filament channel are expected to be related to the evolution of the photospheric magnetic field.
Then, our analysis first focuses on the appearance of two bipoles on both sides of the filament channel in HMI maps (Figure~\ref{HMI_time} d, yellow and cyan boxes). 
The HMI movie allows us to follow their time evolution.

Figure~\ref{HMI_time} (panel d) shows the bipole n$_3$ p$_3$ (yellow box) and its magnetic flux evolution (panel e). The proximity of the two polarities gives the impression of an emerging flux. Moreover, it is associated with an EUV bright point (Figure~\ref{AIA_HMI}, panels d, e). However, this bipole does not correspond to an emerging bipole but to the transport of a negative magnetic polarity that emerged earlier within the filament channel.  %First 
In the movie, we observe that small polarities form within supergranules in the channel. Next, they rapidly move to the supergranule borders, where small-scale cancellations change their shapes and fluxes. Then, the tiny negative polarities forming n$_3$ rapidly cross the PIL, gather near p$_3$, which is also moving quickly, and enter the box around 08 UT on September 29. 
To better follow this evolution, we perform a time distance plot (Figure~\ref{hmi_ts} b).
Finally, the polarity n$_3$ progressively cancels p$_3$ by the end of September 29, and we note that, already on September 28 before midnight, there is a slight cancellation. 
In a very similar study of filament formation \citep{Schmieder2014}, the granules were used to compute the displacement of the parasitic polarities. Supergranules were identified, and the parasitic polarities of the feet were found to be located at the borders of supergranules.  The same process was also noted in \citet{Li2022}.

On the other side of the filament, another bright point corresponds to the magnetic polarities n$_4$ p$_4$ (Figure~\ref{HMI_time} d, f). At first glance, it appears to correspond to the emergence of a bipole; however, it is also due to the transport of negative polarity toward positive polarity. The bipole is formed by the convergence of n$_4$ towards p$_4$. The positive polarity flux p$_4$ is mainly increasing on September 28, while the negative polarity is moving to the west at a velocity $\approx 0.3$ \kms\ on September 29 (Figure~\ref{hmi_ts} c).  Both polarities enter the cyan box (Figure~\ref{HMI_time} d) and are approaching each other, but the positive and negative fluxes only weakly cancel (Figure~\ref{hmi_ts} c).  

\begin{figure} %%%%%%%%%%%%%%%%%% FIGURE 8
    \centering
    \includegraphics[width=1\textwidth]{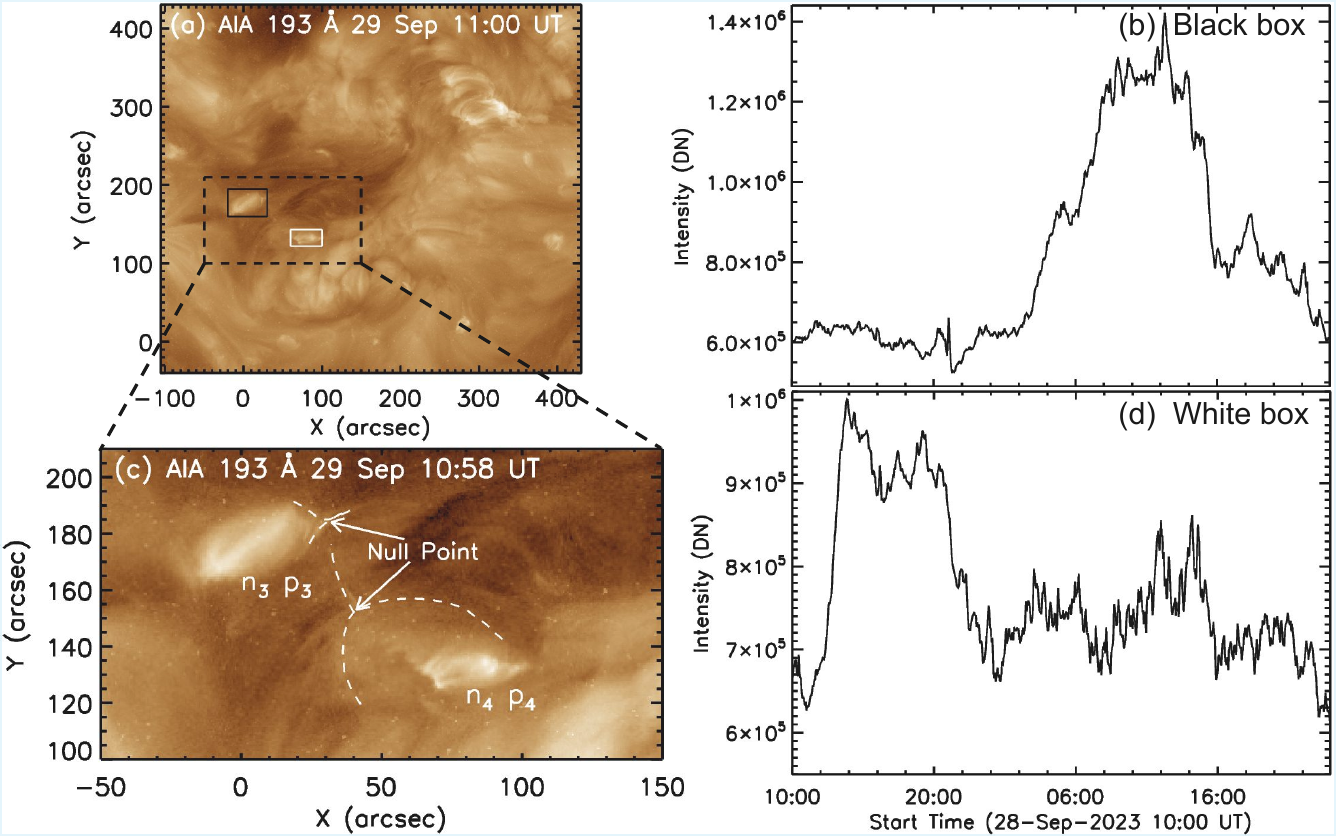}
    \caption{Panel (a) shows the filament observed in AIA 193 \AA. The black dashed box shows the FOV of panel (c). The black and white small solid box shows the regions in which the intensity is plotted with time (panels (b) and (d)). 
    Panel (c) shows brightenings observed in AIA 193 \AA\ corresponding to the two bipoles n$_3$ p$_3$ and n$_4$ p$_4$. The white arrows show the possible magnetic null points above the two bipoles. The loop-like connection between the nulls, plausibly a plasma trace of the separator linking them, is outlined in a dashed line. Other dashed lines indicate intensity limits that plausibly mark separatrices.
    Panels (b) and (d) represent the intensity variation within the black and white boxes shown in panel (a), respectively.
    }
    \label{null_point}
\end{figure}

In the AIA 193 \AA\ movie, attached to Figure~\ref{AIA_HMI}, we note the extension evolution of the coronal arcades above the bipoles n$_3$ p$_3$ and n$_4$ p$_4$, which is expected to be linked to the observed filament located in between.  Figure~\ref{null_point} c displays a zoom of the possible link between the two bipoles, where {the elongated and corner-like shape of the plasma emission indicates that magnetic null points are plausibly present. There is even the presence of a loop-like emission in between the two bipolar regions (outlined by a white dashed line). We interpret it as the separator linking the two null points. A separator is the intersection of two separatrices which are locally the fan planes of the null points. The trace of these separatrices, separating plasma regions with different magnetic connectivities, is outlined by an abrupt change of AIA 193 emission. These separatrice traces are outlined by white dashed lines in Figure~\ref{null_point} c.   
Finally, the plasma trace of the separator partially disappears before the filament reforms, suggesting a plausible local magnetic-field reconfiguration if the evolution is not purely thermal.  Finally, the emission evolution of the coronal arcades above the bipoles n$_3$ p$_3$ and n$_4$ p$_4$, shown in Figure~\ref{null_point} b, d, will be described within the next subsection in relationship with the filament oscillations. } 

Next, we study the magnetic field present in the filament channel.
The HMI movie presents the evolution of magnetic flux in the filament channel (see the attached movie). Tiny emerging polarities develop inside supergranules and move across the PIL.
The two former studied bipoles are examples, but there are many other tiny bipoles with very low magnetic flux. 
It is difficult to track changes in supergranules, and even more challenging to monitor the emergence of tiny polarities within them. These changes in the movement and cancellation of magnetic flux could help the filament channel extend to F$_3$.  Similarly, \citet{Li2022} explains the extension of their studied filament by successively cancelling flux below the filament. This supports the model of \citet{Martens1989}.

\begin{figure}[!h]    %%%%%%%%%%%%%%%%%% FIGURE 9
\centerline{\includegraphics[width=1.\textwidth,clip=]{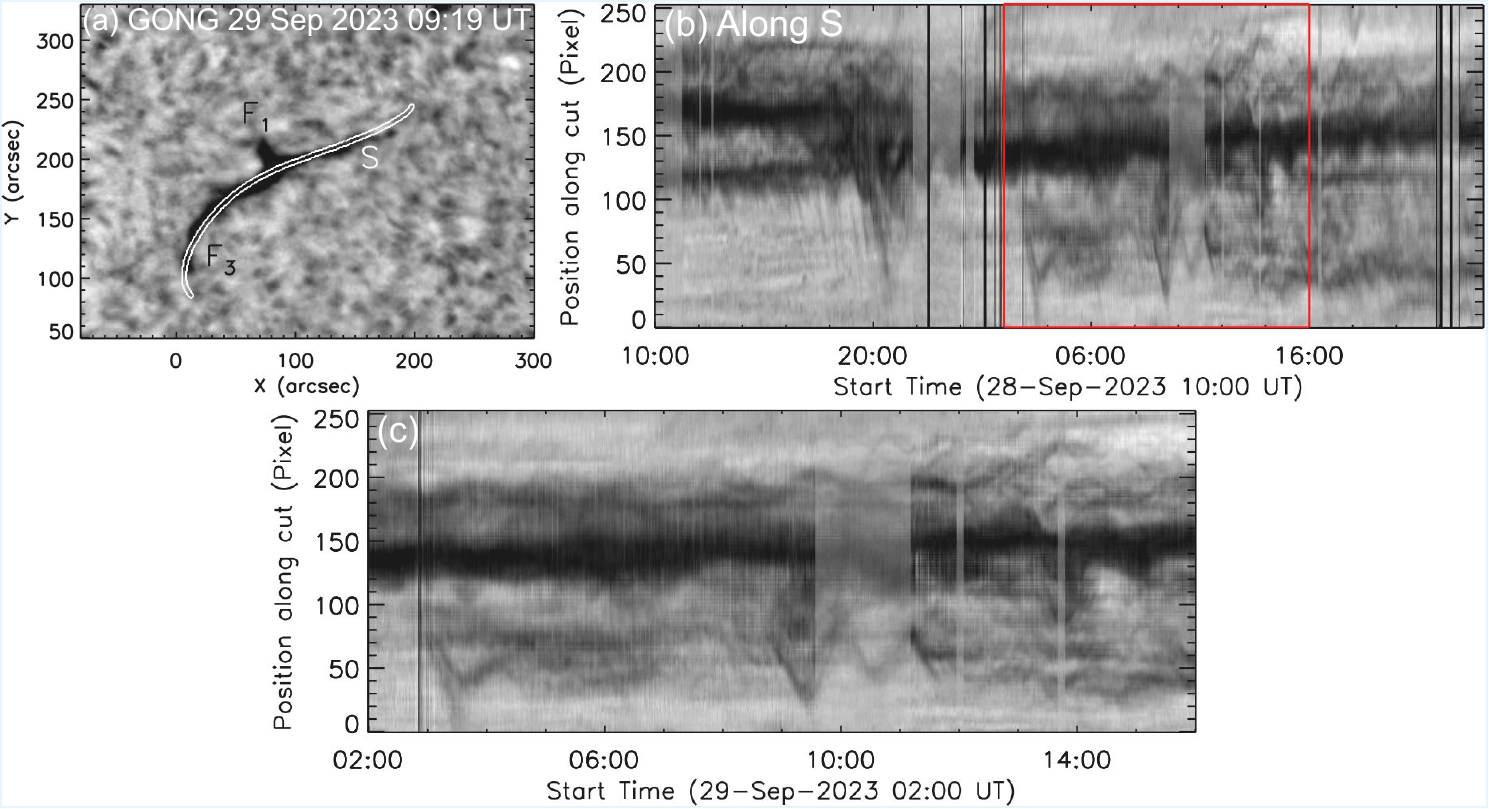}}
      \caption{ Panel (a) presents the filament observed by GONG  on September 29. The white curve S is the slit used to detect the filament motions. The slit is oriented from south-east to north-west. F$_1$  is a footpoint of the filament, and F$_3$ is the extension to the south. 
      The slit width is 2 pixels. Panel (b) shows the time-distance diagram along the slit S. The time-distance diagram enclosed within the red box (from 02:00 to 16:00 UT on September 29) is shown in panel c.
      }
\label{timeslice}
\end{figure}

\subsection{Longitudinal Amplitude Oscillations} 
\label{LAO}
We have observed oscillations in the filament along a slit parallel to the main filament axis. 
Therefore, the observed oscillations are almost longitudinal in nature. These longitudinal oscillations along the filament spine are observed over two days using GONG data. The data for both days have been aligned to a common reference time, as described in Section~\ref{Instrument}.  To analyze the longitudinal amplitude oscillations along the filament, we employ the time-distance method using a curved slice S, oriented from southeast to northwest, to trace the plasma motions along the filament spine (Figure~\ref{timeslice} a). Figure~\ref{timeslice} b presents the time-distance plot for the filament observed from September 28 at 10:00 UT until September 29 at 23:59 UT, during which the extension F$_3$ becomes visible. Figure~\ref{timeslice} c is a zoom during the time period with more/larger oscillations.
The gaps present in the data are indicated by the black regions in the time–distance plot. The filament structure undergoes significant changes during this time period (see the accompanying GONG movie). 

Longitudinal oscillations are detected at both ends of the filament on September 29, where the contrast with the surroundings is higher.  In the center of the filament, the oscillations seem to be mixed because of a larger number of fibrils. Moreover, the oscillations may not be exactly along the main axis of the filament, but rather along the fine structures of the filament, which make an angle of approximately 20 degrees (see section~\ref{evolution}).
On September 29, a few oscillatory cycles were detected near F$_3$ between 08:00 and 12:00 UT (Figure~\ref{timeslice} b).  These oscillations are preceded by two tentative extensions around 21:00 UT on September 28 and around 03:00 UT on September 29.
The oscillation period is approximately 70 min. 
It is a common period for filament longitudinal oscillations \citep{Luna2018, Luna2022}. 
Moreover, several threads along the filament spine also display oscillatory behavior in the time–distance diagrams on both days. 

The longitudinal amplitude oscillations along F$_3$ reflect its dynamical evolution.  
On September 29, after 12 UT, parts of the extension F$_3$ are relatively stable. 
The filament magnetic structure with a long FR may have tentatively existed previously (e.g, around 20:23 UT on September 28). Photospheric motions lead to emerging magnetic flux, which then cancels, and heating as a consequence of reconnection. This process supplies energy into the overlying magnetic configuration and the plasma embedded within it. In Figure~\ref{null_point} b, the intensity curve over n$_3$ p$_3$ shows an enhancement at approximately the time of the longitudinal oscillations in F$_3$ between 08:00 UT and 12:00 UT.  The evaporation of the hot plasma during reconnection could trigger longitudinal oscillations, pushing the cold plasma along the magnetic field lines. This initiates a pendulum mode of the cool plasma that fills a larger portion of the magnetic dips \citep{Luna2012,Luna2018, Zhang2017}. 
Such a relationship is not clear with the earlier brightness evolution over n$_4$ p$_4$ (Figure~\ref{null_point} d) as F$_3$ was not well formed at that time.

\begin{figure}    %%%%%%%%%%%%%%%%%% FIGURE 10
\centerline{\includegraphics[width=1.0\textwidth,clip=]{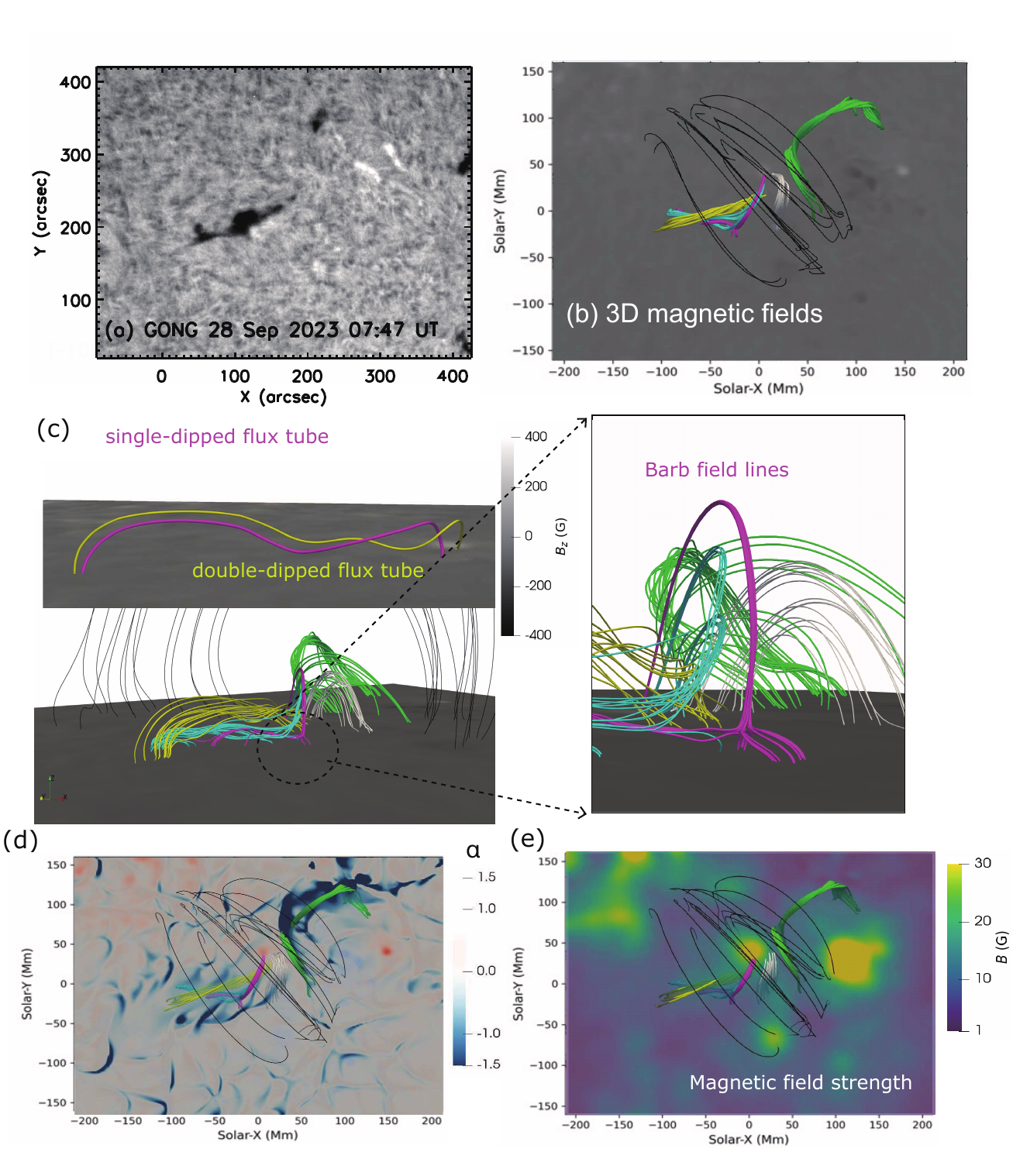}}
       \caption{Filament observation on September 28 by GONG  in  helioprojective coordinates (a) and  3D  reconstruction of the %computed 
       coronal magnetic structures  in the local coordinate system after projection correction (b--e). %3D computed coronal magnetic structures (b--e) and the comparisons with GONG observations (a). 
       The yellow and green field lines show the southern and northern parts of the filament spines, respectively. The cyan and magenta lines represent the field lines hosting filament feet F$_1$ and F$_2$ in positive and negative main polarities, respectively. The zoomed views in panel (c) show field lines associated with the feet as viewed from the side. Panels (d) and (e) display, at $z= 20$ Mm, the distribution of twist density ($J\cdot B/B^{2}$) and magnetic-field strength, respectively, with overlaid the magnetic field lines in yellow, green, magenta and cyan. The green field lines in the extrapolation correspond to the filament material in the northern part, where continuously distributed magnetic dips are lacking and thus cannot support a long, coherent filament as in the southern part.}
\label{FR_28}
\end{figure}

\begin{figure}    %%%%%%%%%%%%%%%%%% FIGURE 11
\centerline{\includegraphics[width=0.8\textwidth,clip=]{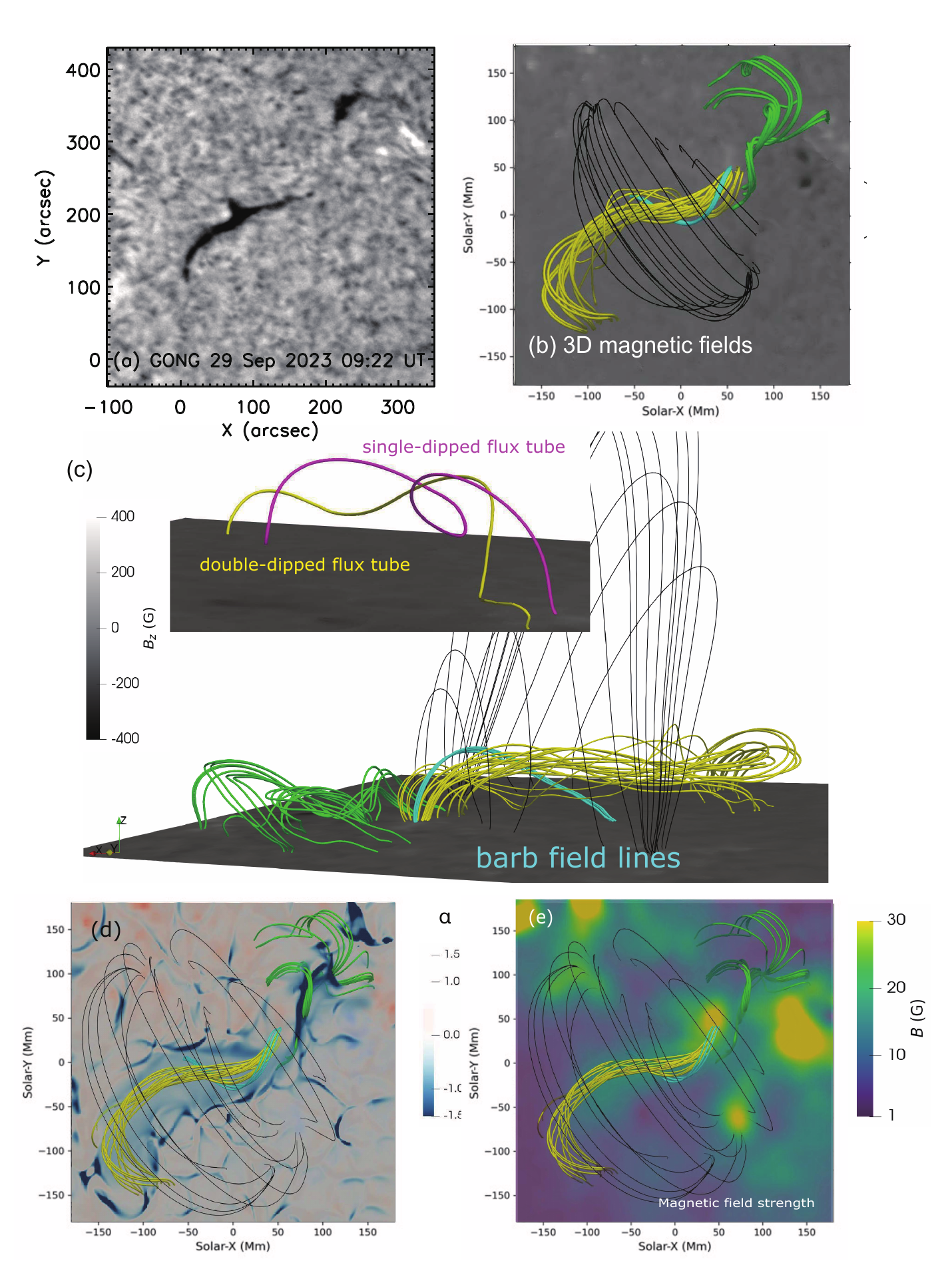}}
       \caption{
       The same as Figure~\ref{FR_28} but for the filament on September 29.}
\label{FR_29}
\end{figure}

\begin{figure}    %%%%%%%%%%%%%%%%%% FIGURE 12
\centerline{\includegraphics[width=1.\textwidth,clip=]{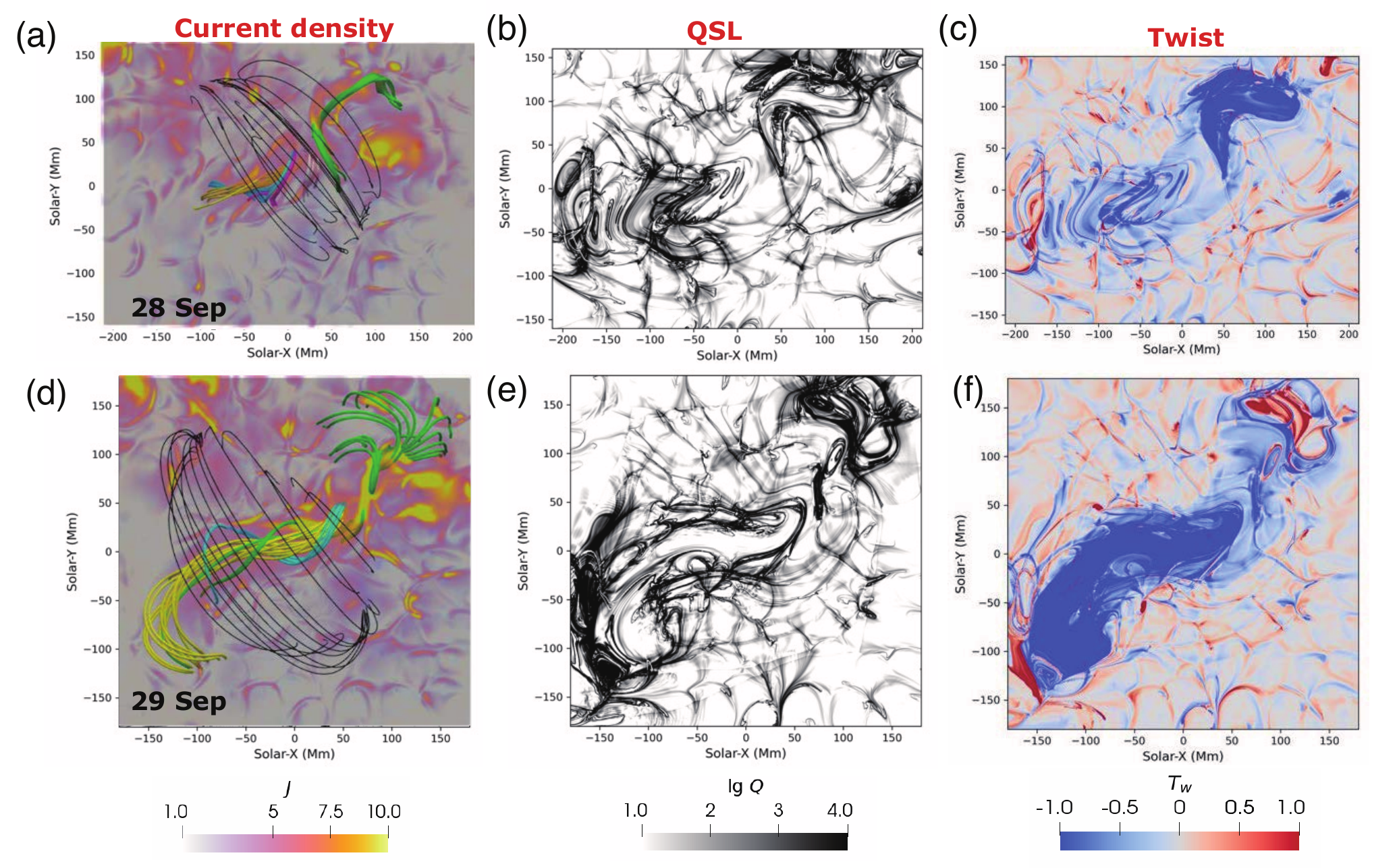}}
       \caption{
       Current density, quasi-separatrix layers (QSL), twist maps computed for the filament on September 28 (panels a, b, c) and for September 29 (panels d, e, f). The unit of the current density is 10$^{-2}$ A m$^{-2}$.
       }
\label{QSL}
\end{figure}

\section{Numerical Construction of 3D magnetic fields of filaments}
\label{simu}

\subsection{Numerical setup}
To investigate the 3D magnetic structures of solar filaments, we reconstruct their 3D magnetic fields with the magneto-frictional method, in which the initial magnetic fields are provided by the potential field extrapolation and the Biot-Savart laws regularized in FR model \citep{Titov2018}. The resulting magnetic field is magneto-frictional relaxed to a non-linear-force-free-field (NLFFF) extrapolation based on  \citet{Guo2016a, Guo2016b}. This is achieved in the MPI-AMRVAC framework \citep{Xia2018, Guo2016a, Guo2016b, Guo2019, Guo2021_twist}.

The implementation process is as follows. First, we correct for projection effects and remove the Lorentz force and torque using an optimisation technique to conform to the force-free assumption \citep{Wiegelmann2006}. Hereafter, we extrapolate the potential magnetic fields using the Green's function based on the vertical $B_{z}$ component of the preprocessed vector magnetogram. Next, we insert a FR with regularised Biot-Savart laws, in which the parameters are constrained by the observed filament shape, as described in \citep{Guo2019} and \citet{Guo2021_twist}. The parabolic profile for the current density is adopted in this model, leading to a FR with a strongly twisted core and a weakly twisted outer shell \citep{Guo2021_twist}.

The FR is mainly controlled by four parameters: the FR path (C), minor radius (a), toroidal flux (F), and electric current (I). Among them, the first two parameters are approximated as the filament path and width, respectively. The toroidal flux is estimated as the average unsigned flux of two flux-rope footpoints, and the electric current is calculated by the equilibrium condition \citep{Titov2018}.   
Then, the combined potential and FR fields are relaxed to a force-free state using the magneto-frictional method. The computational domain of the model on September 28 and 29 is ${[x_{min},x_{max}]\times[y_{min},y_{max}]\times[z_{min},z_{max}]=[-180,180]\times[-180,180]\times[1,307]}$ Mm$^{3}$ and ${[x_{min},x_{max}]\times[y_{min},y_{max}]\times[z_{min},z_{max}]=[-211,211]\times[-160,160]\times[1,322]}$ Mm$^{3}$, 
respectively.

\subsection{Results analysis}

Figures~\ref{FR_28} and \ref{FR_29} display the 3D magnetic structures of the filament on September 28 and 29, along with comparisons to GONG $\rm H{\alpha}$ observations, respectively. Both observations and NLFFF extrapolation show that the filament consists of two separate parts (yellow and green field lines). In the previous section, only the southern part was analyzed. Here, we consider both sections, north and south. 

On September 28, the southern part consists of a twisted FR with deep magnetic dips, while the northern part corresponds to a sheared arcade with shallow dips (Figure~\ref{FR_28}).  According to \citet{Guo2022_prom} and \citet{Guo2021_2Dip}, stationary filament threads are more prone to remain in deep dips with FRs, whereas dynamic threads formed due to the thermal non-equilibrium cycle favor weakly twisted FRs and sheared arcades. The magnetic structure on September 29 is still composed of two parts, while the southern FR (yellow lines) is more extended (Figure~\ref{FR_29}). Some field lines of the southern FR are nearly joining the northern positive parasitic polarities, while those in the south connect to the southern negative parasitic polarities.  Our numerical model can explain the following observational phenomena. 

First, the magnetic structure of the studied filament is a FR  composed of two separate parts, deviating from a coherent structure, which can explain the two portions of the filament in observations (as shown in Figures~\ref{FR_28} a and~\ref{FR_29} a). The magnetic configuration in the southern segment corresponds to an FR,  while that in the north is a sheared arcade, producing a greater number of magnetic dips that can host more filament plasma. 

Second, regarding the formation mechanism of filament feet, it is found that the foot F$_2$ is due to the positive parasitic magnetic polarity p$_2$ embedded in negative polarities (magenta lines in Figure~\ref{FR_28}). 
The zoomed image of Figure~\ref{FR_28} c shows, with magenta field lines, the fan-spine structure associated with a magnetic null point. It is due to the existence of p$_2$ nearby. 
Next, the foot F$_1$ corresponds to twisted field lines (cyan lines) located below the spine field lines (yellow lines).
For the filament on September 29, the foot F$_2$ disappears with the vanishing of the parasitic polarity p$_2$. However, the foot F$_1$ becomes more developed with the growth of the FR.

Third, the comparison between the 3D magnetic structures on September 28 and 29 reveals the filament extension F$_3${ with more extended south-eastward twisted yellow field lines in Figure~\ref{FR_29} than in Figure~\ref{FR_28}. This explains the increase in the filament extension observed by GONG (panels (a) of the same figures).} F$_3$ is located within the yellow field lines (Figure~\ref{FR_29}). The modeled FR is also consistent with the numerous filament threads along the filament spine detected in high-resolution observations (Figure~\ref{THEMIS_IRIS}). 

Fourth, in Figures~\ref{FR_28} d and~\ref{FR_29} d, the twist density is shown at $z= 20$ Mm. The twist density is defined as $\alpha = \vec{J} \cdot \vec{B}/B^2$, which indicates how much the field lines are non-potential. 
We note a significant change between the two days with the formation of a large channel {with low $\alpha$ values around the field lines of the FR (yellow lines). This channel is bordered by intense $\alpha$ values distributed along linear structures, especially on September 29. }
  
Finally, in Figure~\ref{QSL} we present the current density, the quasi-separatrix layers (QSLs) and the twist number for the two days. Quasi-separatrix layers (QSLs) represent regions where magnetic connectivity changes rapidly \citep{Priest1995, Demo1996} and are often used to delineate the boundaries of flux ropes.
The twist number is the number of turns along field lines (so a curvilinear integration of $\alpha$). This is a nondimensional parameter, and, in particular, it is independent of $B$ magnitude.  
The twist is weak on September 28 (except in the northern region), then strengthens and extends on September 29. This is traced indirectly by the filament extending southward.
Next, at $z=20$ Mm, most small-scale connectivities have been filtered out, allowing us to observe the large-scale topology with QSLs. They separate the twisted/sheared core of the filament B configuration from the surrounding, more potential arcade (where flux cancellation has not yet been active because it is too far from the PIL). 
The current density maps show a current layer separating the core from the surroundings, as QSLs do. So on both days, both twist and QSL at $z= 20\;$Mm are consistent. Such current concentrations on QSLs, derived from magnetic extrapolation, were previously reported \citep{Guo2013}. With QSLs present, such current layers are expected to form during MHD evolution; here, however, this is inferred from two magnetic extrapolations.  This is a confirmation that the extrapolations are capturing well the key aspects of the magnetic configuration evolution.  In summary, with these maps for the two days, we can clearly characterize the buildup of the FR between 28 and 29 September with the twist maps, and the separation of the FR from the surrounding more potential fields with the current density and the QSLs.

\section{Conclusion} 
\label{con}

Coordinated observations were obtained for two days during a THEMIS campaign in September 2023, involving the Hinode/SOT and IRIS spacecraft. The quiescent filament was also observed by GONG, exhibiting high dynamics during its elongated formation. 
In addition, THEMIS observations of the filament show multiple strands in its spine and feet. The long-term study shows large-amplitude oscillations in the filament's extension. The evolution of the photospheric magnetic field is well observed with the HINODE/SOT-SP and HMI instruments. Magnetic polarities emerge, then some of them cross the PIL, inducing magnetic cancellation, a key ingredient for FR formation.

Magnetic extrapolations are obtained by inserting an FR into a potential field; then, a magneto-frictional method is applied.  
The deduced magnetic configurations include two separate regions having magnetic dips. They are located where the two filament parts are observed (along the same PIL). 
The FR extends in length between the two days, as the filament does in the observations. This changing shape is due to the emergence of magnetic flux, which then cancels magnetic polarities.  

The observations and magnetic extrapolations provide key insights into the formation and extension of the filament:
\begin{itemize}
 \item The formation mechanism of filament feet is consistent with the dip model \citep{Aulanier1998_D}. Dips are due to field lines being ``attracted" by magnetic parasitic polarities with opposite signs to the background.  The feet evolve with the parasitic polarities.
\item The filament extension is formed progressively by canceling flux along the PIL. This progressively transforms the sheared magnetic field into an FR. 
\item Large amplitude oscillations are observed to develop in a fraction of time during the two days. 
In the filament channel, cancellation and polarity motions impose an evolution on the magnetic polarities.  Therefore, the magnetic dips change shape and location, which forces the dense plasma to evolve.  
Moreover, heating due to magnetic reconnection of a bipole at the channel boundary is plausibly an energy source for the oscillations.  The cold plasma is pushed by the hot plasma, initiating the oscillations. Finally, similar periods of around 70 min for the longitudinal oscillations are observed across different dips. This is in agreement with theory \citep{Zhou2017}.
\item The two magnetic bipoles, formed by converging flows at the border of the filament channel, organize the magnetic field lines in the extension of the filament, which develops on the second day.
\item  We analyze the changes in the magnetic configuration between the two days using 3D MHD reconstruction with two magnetograms. On the 28$^{th}$ of September, the FR begins to form, and the next day it is more extended, in agreement with filament observations. 
On the 29$^{th}$ of September, the FR is located in a channel bordered by QSLs with high electric currents.
They are separating the FR from the nearly surrounding potential field.

\item The magnetic extrapolation reveals double-dipped flux tubes, supporting the scenario of magnetically connected threads, as shown in Figures~\ref{FR_28} c and \ref{FR_29} c. The second dip corresponds to the extension of the filament in F$_3$.
\end{itemize}

The presence of multiple threads along one flux tube introduces new physics. For example, the longitudinal oscillations of the threads are no longer independent, and there is thread-to-thread interaction, which significantly changes the damping time of the oscillations, sometimes leading to decayless oscillations \citep{Zhou2017}.
The signatures of such thread-thread interactions were observed previously \citep{Zhang2017}.

In conclusion, these observations and the 3D MHD modeling confirm the importance of the photospheric magnetic field within the filament channel. The magnetic field spatial distribution and temporal evolution structures the magnetic configuration of the filament, and in particular the FR. 
On September 28 the filament is formed by a split flux tube. One part of the flux tube is rooted in the photosphere aside an observed interruption in the filament. 
Then, the combination of high resolution observations and numerical modeling is an efficient way to understand the physics present behind the evolution of filaments.
The dense and cold plasma caught in the magnetic configuration provides important constraints for the modeling the magnetic configuration evolution. 
Parasitic polarities, associated with filament feet, result in secondary dips above the related local inversion line. These dips belong to long field lines that pass below the flux tube. Many of these field lines are not rooted near the related foot. 

\begin{acknowledgments}
The authors thank the referee for constructive comments and suggestions, which significantly improved the manuscript. This work is based on ground-based observations obtained by the THEMIS telescope in Tenerife in the Canary Islands, operated by Bernard Gelly, Richard Douet, and Didier Laforgue$^\dagger$ during a multi-wavelength campaign with IRIS (IHOP444 - PIs Nicolas Labrosse and Brigitte Schmieder - coordinated with Hinode by Sarah Matthews). Hinode is a Japanese mission developed and launched by ISAS/JAXA, with NAOJ as a domestic partner and NASA and STFC (UK) as international partners. It is operated by these agencies in co-operation with ESA and NSC (Norway).
The  H$\alpha$ spectroheliograph was provided by BASS2000.obspm.fr. AIA data are courtesy of NASA/SDO and the AIA, EVE, and HMI science teams. IRIS is a NASA small explorer mission developed and operated by LMSAL, with mission operations executed at NASA Ames Research Center and major contributions to downlink communications funded by ESA and the Norwegian Space Centre. B.S. thanks Rony Keppens and Tom Van Doorsselaere for fruitful discussions on waves and filament fine structures.
G.K. acknowledges support from DST INSPIRE.
J.H.G.\ is supported by the fellowship of the China National Postdoctoral Program for Innovative Talents under Grant Number BX20240159. 
S.P.\ is funded by the European Union (ERC, Open SESAME, 101141362). Views and opinions expressed are, however, those of the author(s) only and do not necessarily reflect those of the European Union or the European Research Council. Neither the European Union nor the granting authority can be held responsible for them. S.P.\ is funded by the projects C16/24/010 (C1 project Internal Funds KU Leuven), G0B5823N and G002523N (WEAVE) (FWO-Vlaanderen), and 4000145223 (SIDC Data Exploitation (SIDEX2), ESA Prodex).

\end{acknowledgments}

\begin{contribution}
G.K.\ provided the formal analysis of the data and validation. She also edited the manuscript.
J.H.G.\ is responsible for the numerical simulations.
B.S.\ and R.C.\ supervised the work and partly wrote the manuscript. P.D.\ edited the manuscript. S.P. revised the manuscript.
B.G.\ was providing the THEMIS observations.
\end{contribution}

\facilities{GONG, IRIS, Meudon spectroheliograph, SDO (AIA and HMI), Hinode/SOT, and THEMIS.}

\appendix
\section{Description of Instruments}
\label{appendix}
The Solar Dynamics Observatory \citep[SDO;][]{Pesnell2012} consists mainly of two instruments: the Atmospheric Imaging Assembly \citep[AIA;][]{Lemen2012} and the Helioseismic and Magnetic Imager \citep[HMI;][]{Schou2012}. The AIA observes the full Sun in seven extreme ultra-violet (EUV; 94, 131, 171, 193, 211, 304, 335 \AA), two ultra-violet (UV; 1600, 1700 \AA) and one white light (4500 \AA) wavebands, with a pixel size and  temporal sampling of 0.6$''$ and 12 s, respectively. Here, we have mainly used AIA 193 and 304 \AA\ filter data to observe the filament. HMI provides the photospheric magnetic field. %and the distribution of sunspots on the Sun. 
It observes the Sun at a wavelength of 6173 \AA\ with a pixel size of 0.5$''$ and a temporal sampling of 45 s. For the present study, we have used longitudinal magnetograms to analyse the magnetic field of the filament region. We have also used HMI vector magnetograms for 3D numerical reconstruction.

The Interface Region Imaging Spectrograph \citep[IRIS;][]{DePontieu2014} provides spectra in three different passbands, 1332 – 1358 \AA\, 1389 – 1407 \AA, and 2783 – 2834 \AA, these passbands include several spectral lines such as Mg II h (2803 \AA) and Mg II k (2796 \AA) formed in the chromosphere, as well as C II (1334/1335 \AA) formed in high chromosphere and Si IV (1394/1403 \AA) formed in the transition region. IRIS also provides slit jaw images at four different UV wavelengths: Mg II wing (2830 \AA), Mg II k (2796 \AA), C II (1330 \AA), and Si IV (1400 \AA). IRIS observed the filament from 10:29:57 UT to 11:22:21 UT, with its FOV centered at $x = - 130''$ and $y = 174''$ in coordination with THEMIS on September 28 and from 10:34:17 UT to  11:26:37  UT with the FOV center at $x=74 ''$ and  $y=175''$  on September 29. 
During one observation hour, 
IRIS acquired 3 rasters with 64 steps, % from 10:29:57 to 10:47:25, and 10:47:42 to 11:04:37 and 11:04:53 to 11:22:21, 
each step, having a size of $2''$ covered a FOV of $127'' \times 175''$ in $\approx 20$ min, with raster cadence of 1048 s and step cadence of 16.4 s. The slit jaw images are taken in the 2796 \AA\ (Mg II k) wavelength with a cadence of 16 s that covers the FOV of $166'' \times 175''$. The SJIs were obtained only in the Mg II k 2796 \AA\ passband.

The Solar Optical Telescope \citep[SOT;][]{Tsuneta2008}, is one of the three instruments onboard Hinode \citep{Kosugi2007}. The SOT is a 50 cm diffraction-limited telescope equipped with both a filtergram and a spectropolarimeter (SP). The SP instrument acquires the spectra of the two Fe lines at 630.15 and 630.25 nm, along with the nearby continuum. SP can operate in four different modes: normal map, fast map, dynamics, and deep magnetograms. For the current study, we have used SP observations on September 28 at 10:59 UT and on September 29 at 10:35 UT. Both SP observations are operated in fast map mode, with a pixel size of 0.32$''$ along the slit and 0.30$''$ transverse to the slit, covering a FOV of 145$'' \times 164''$ on September 28 and a FOV of 153$''\times 164''$ on September 29. The spectral scale is $\approx 21.5\;$m\AA. In this mode, a region of $1.6''$ wide is covered in 18\;s.

The Global Oscillation Network Group \citep[GONG;][]{Harvey1996} is a network of six identical telescopes located worldwide that provides full-disk images of the Sun in H$\alpha$ wavelength with a pixel size of 2.5$''$ and a temporal resolution of 1~min. We have used the GONG H$\alpha$ data to analyse the filament.

The T\'elescope H\'eliographique pour l’Etude du Magn\'etisme et des Instabilit\'es Solaires, \citep[THEMIS;][]{Mein1985} built in the 90's and recently renovated with adaptive optics  \citep{Schmieder2025}, allows observations in spectropolarimetric mode, a small region of the Sun ($ 80'' \times 120'' $) with a pixel size of 0.06$''$ along the slit and a spectral dispersion of $\approx$ 3.067 m \AA\ per pixel. THEMIS provides spectra from four cameras at different wavelengths. Here, we focus on the H$\alpha$ observations with a passband centred at 6563 \AA\ and 6.3 \AA \ wide. The slit step is either 0.5$''$ or 1$''$, with a slit width of 0.5$''$. Depending on the targets and viewing conditions, the H$\alpha$ exposure time ranges from 0.05 to 0.2 s. 

%It observes the full Sun with a {\bf a spatial sampling of 1.1$''$} and a temporal resolution of 1 minute, respectively. %*** I am not convinced of the relevance of this, especially at the end: *** This conclusion corroborates the new view of  the formation of filament feet with minima of magnetic field strength between supergranules, and  no direct evidence of  parasitic polarities \citep{Chen2025}.} %*** Does the following answer the referee questions? I am not convinced. The referee looks more interested in more general questions, just as the beginning of this paragraph. *** This split topology is due to strong network polarities on the edge of the filament channel as it was  shown in \citet{Dudik2008}. The former authors studied  the topological departures from translational invariance along a filament. They calculated the coronal magnetic field from a ``linear magnetohydrostatic'' extrapolation of a composite magnetogram and detailed the shape of the dips corresponding to the long observed filament. They showed the importance of the network at the border of the filament. 

\bibliography{references}{}
\bibliographystyle{aasjournalv7}

\end{document}